\documentclass[conference]{IEEEtran}
\IEEEoverridecommandlockouts
\usepackage{cite}
\usepackage{amsmath,amssymb,amsfonts}
\usepackage{graphicx}
\usepackage{textcomp}
\usepackage{xcolor}
\usepackage{color}
\usepackage{colortbl}
\usepackage{float}
\usepackage{graphicx}
\usepackage{subcaption}

\usepackage{algorithm}
\usepackage{algpseudocode}
\usepackage{mwe}

\makeatletter 
\newenvironment{breakablealgorithm}
  {
   \begin{center}
     \refstepcounter{algorithm}
     \hrule height.8pt depth0pt \kern2pt
     \renewcommand{\caption}[2][\relax]{
       {\raggedright\textbf{\ALG@name~\thealgorithm} ##2\par}%
       \ifx\relax##1\relax 
         \addcontentsline{loa}{algorithm}{\protect\numberline{\thealgorithm}##2}%
       \else 
         \addcontentsline{loa}{algorithm}{\protect\numberline{\thealgorithm}##1}%
       \fi
       \kern2pt\hrule\kern2pt
     }
  }{
     \kern2pt\hrule\relax
   \end{center}
  }
\makeatother

\newcommand{\xoverbrace}[2][\vphantom{A}]{\overbrace{#1#2}}
\newcommand{\headersty}[1]{{\normalfont\normalsize\centering\scshape #1}}
\newcommand{\unaryminus}{\scalebox{0.75}[1.0]{\( - \)}}
\newcommand{\algrule}[1][.2pt]{\par\vskip.5\baselineskip\hrule height #1\par\vskip.5\baselineskip}

\def\BibTeX{{\rm B\kern-.05em{\sc i\kern-.025em b}\kern-.08em
    T\kern-.1667em\lower.7ex\hbox{E}\kern-.125emX}}

\title{Deep Positron: A Deep Neural Network Using the Posit Number System}

\author{
\IEEEauthorblockN{Zachariah Carmichael$^{\S}$, Hamed F. Langroudi$^{\S}$, Char Khazanov$^{\S}$, Jeffrey Lillie$^{\S}$, \\ John L. Gustafson$^{\ast}$, Dhireesha Kudithipudi$^{\S}$}
\IEEEauthorblockA{$^{\S}$Neuromorphic AI Lab, Rochester Institute of Technology, NY, USA}
\IEEEauthorblockA{$^{\ast}$National University of Singapore, Singapore}\vspace{-7.2mm}}

\begin{document}
\bstctlcite{IEEEexample:BSTcontrol}
\maketitle


\begin{abstract}

The recent surge of interest in Deep Neural Networks (DNNs) has led to increasingly complex networks that tax computational and memory resources. Many DNNs presently use 16-bit or 32-bit floating point operations. Significant performance and power gains can be obtained when DNN accelerators support low-precision numerical formats. Despite considerable research, there is still a knowledge gap on how low-precision operations can be realized for both DNN training and inference. In this work, we propose a DNN architecture, Deep Positron, with posit numerical format operating successfully at $\leq$8 bits for inference. We propose a precision-adaptable FPGA soft core for exact multiply-and-accumulate for uniform comparison across three numerical formats, fixed, floating-point and posit. Preliminary results demonstrate that 8-bit posit has better accuracy than 8-bit fixed or floating-point for three different low-dimensional datasets. Moreover, the accuracy is comparable to 32-bit floating-point on a Xilinx Virtex-7 FPGA device. The trade-offs between DNN performance and hardware resources, \textit{i.e.} latency, power, and resource utilization, show that posit outperforms in accuracy and latency at 8-bit and below.





\end{abstract}


\section{Introduction}

Deep neural networks are highly parallel workloads which require massive computational resources for training and often utilize customized accelerators such as Google's Tensor Processing Unit (TPU) to improve the latency, or reconfigurable devices like FPGAs to mitigate power bottlenecks, or targeted ASIC’s such as Intel's Nervana to optimize the overall performance. The training cost of DNNs is attributed to the massive number of primitives known as multiply-and-accumulate operations that compute the weighted sums of the neurons' inputs. To alleviate this challenge, techniques such as sparse connectivity and low-precision arithmetic~\cite{DeepCompress2016,wu2018training, Microsoft2018}
~are extensively studied. For example, performing AlexNet inference on Cifar-10 dataset using 8-bit fixed-point format has shown 6$\times$ improvement in energy consumption \cite{hashemi2017understanding} over the 32-bit fixed-point. On the other hand, using 32-bit precision for an outrageously large neural network, such as LSTM with mixture of experts \cite{shazeer2017outrageously}, will approximately require 137 billion parameters. When performing a machine translation task with this network, it translates to an untenable DRAM memory access power of 128~W\footnote{Estimated Power = (20 Hz $\times$ 10 G $\times$ 640 pJ (for a 32-bit DRAM access\cite{DeepCompress2016})) at 45nm technology node}. For deploying DNN algorithms on the end-device (\textit{e.g.} AI on the edge, IoT), these resource constraints are prohibitive.

\begin{figure}
  \centering
  \includegraphics[width=.78\linewidth]{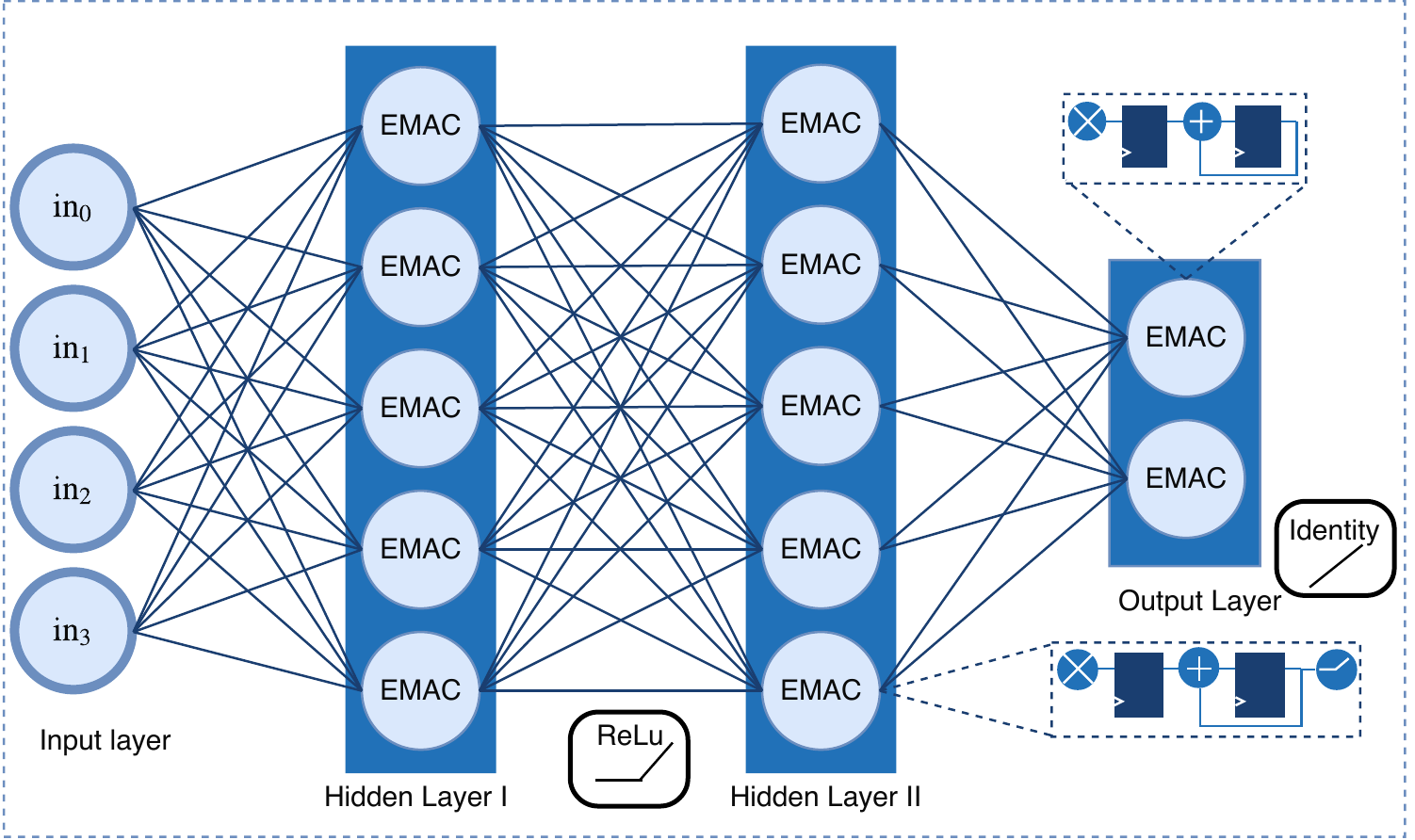}
  \caption{An overview of a simple Deep Positron architecture embedded with the exact multiply-and-accumulate blocks (EMACs).}
  \label{fig:deep_positron_arch}
  \vspace{-5mm}
\end{figure}

Researchers have offset these constraints to some degree by using low-precision techniques. Linear and nonlinear quantization have been successfully applied during DNN inference on 8-bit fixed-point or 8-bit floating point accelerators and the performance is on par with 32-bit floating point \cite{TPU2017,Minerva2016,Microsoft2018}. However, when using quantization to perform DNN inference with ultra-low bit precision ($\leq$8-bits), the network needs to be retrained or the number of hyper-parameters should be significantly increased \cite{mishra2018wrpn}, leading to a surge in computational complexity. One solution is to utilize a low-precision numerical format (fixed-point, floating point, or posit \cite{Ristretto}) for both DNN training and inference instead of quantization. Earlier studies have compared DNN inference with low-precision (\textit{e.g.} 8-bit) to a floating point high-precision (\textit{e.g.} 32-bit) \cite{hashemi2017understanding}. The utility of these studies is limited -- the comparisons are across numerical formats with different bit widths and do not provide a fair understanding of the overall system efficiency.

\begin{figure}
    \centering
    \begin{subfigure}[b]{0.46\linewidth}
        \includegraphics[width=\textwidth]{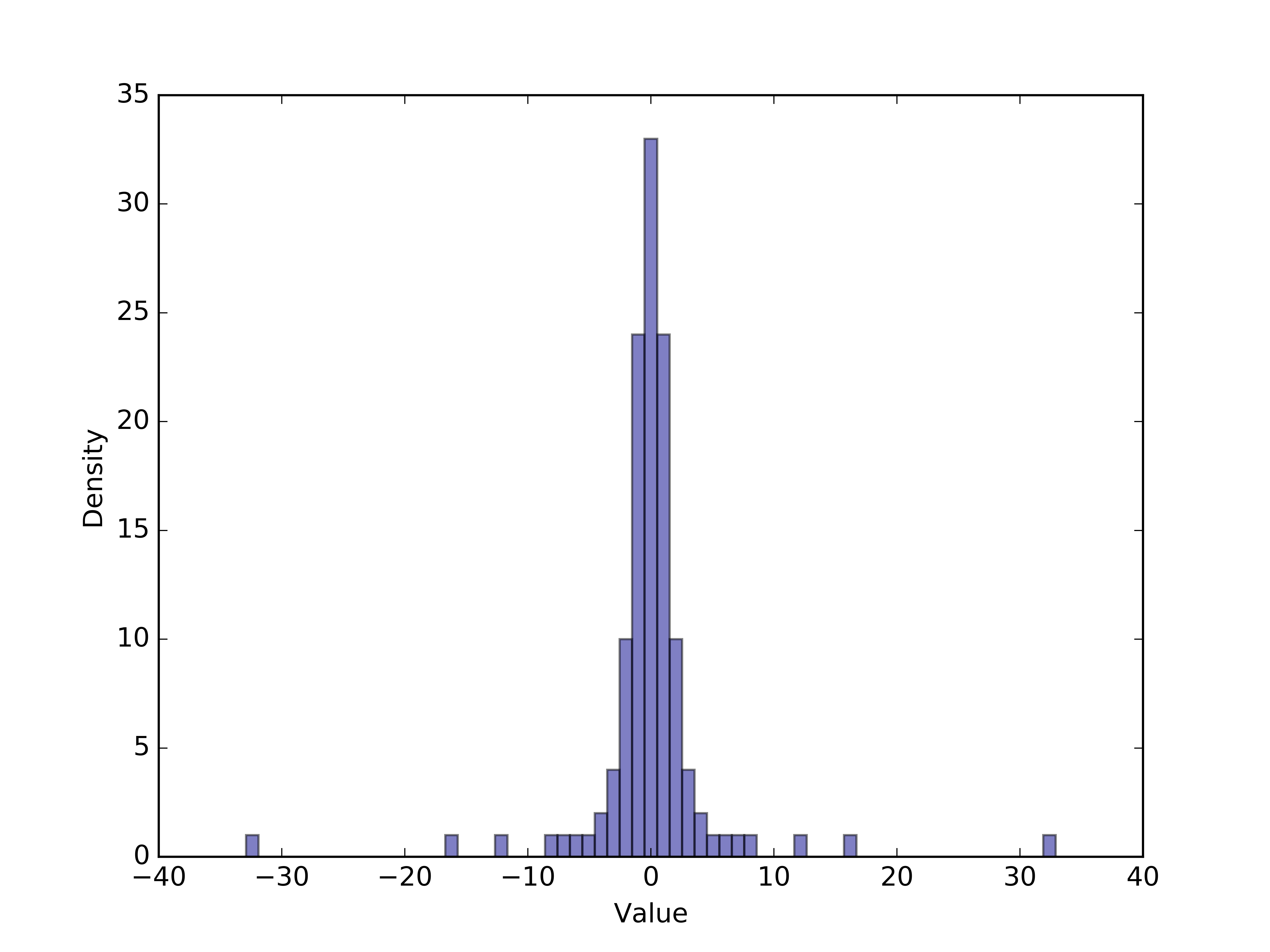}
        \label{fig:gull}
    \end{subfigure}
    \begin{subfigure}[b]{0.46\linewidth}
        \includegraphics[width=\textwidth]{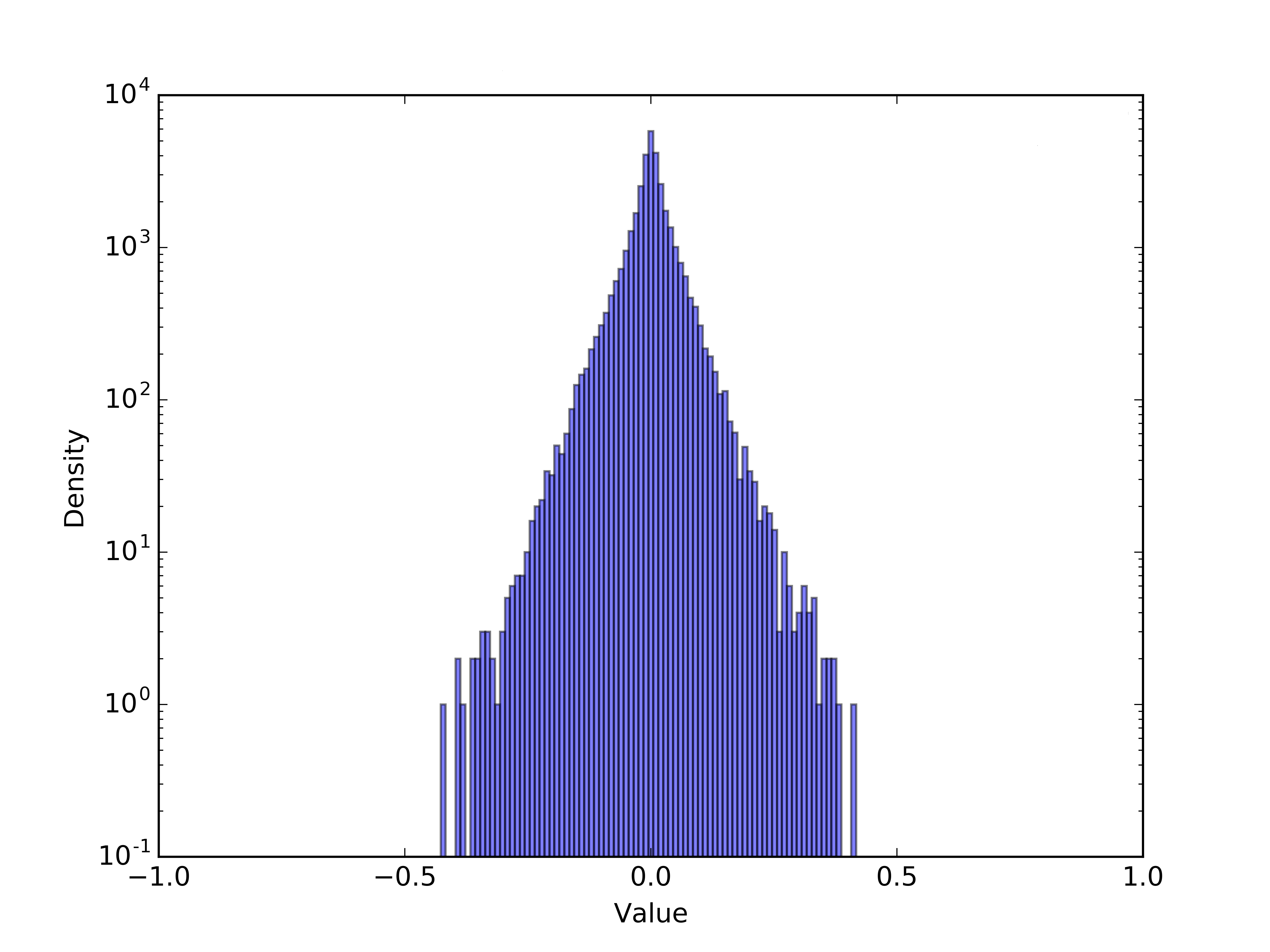}
        \label{fig:tiger}
    \end{subfigure}
    \vspace{-4mm}
    \caption{(a) 7-bit posit ($es=0$) and (b) AlexNet weight distributions. Both show heavy clustering in [-1,1] range.}
    \label{fig:distrubution}
    \vspace{-3.5mm}
\end{figure}

More recently, the posit format has shown promise over floating point with larger dynamic range, higher accuracy, and better closure~\cite{gustafson2017beating}. The goal of this work is to study the efficacy of the posit numerical format for DNN inference. An analysis of the histogram of weight distributions in an AlexNet DNN and a 7-bit posit (Fig.~\ref{fig:distrubution}) shows that posits can be an optimal representation of weights and activations. We compare the proposed designs with multiple metrics related to performance and resource utilization: accuracy, LUT utilization, dynamic range of the numerical formats, maximum operating frequency, inference time, power consumption, and energy-delay-product.



This paper makes the following contributions:
\begin{itemize}
\item We propose an exact multiply and accumulate (EMAC) algorithm for accelerating ultra-low precision ($\leq$8-bit) DNNs with the posit numerical format. We compare EMACs for three numerical formats, posit, fixed-point, and floating point, in sub 8-bit precision.

\item We propose the Deep Positron architecture that employs the EMACs and study the resource utilization and energy-delay-product.

\item We show preliminary results that posit is a natural fit for sub 8-bit precision DNN inference.

\item We conduct experiments on the Deep Positron architecture for multiple low-dimensional datasets and show that 8-bit posits achieve better performance than 8-bit fixed or floating point and similar accuracies as the 32-bit floating point counterparts.

\end{itemize}

\section{Background}
\subsection{Deep Neural Networks}
Deep neural networks are biologically-inspired predictive models that learn features and relationships from a corpus of examples. The topology of these networks is a sequence of layers, each containing a set of simulated neurons. A neuron computes a weighted sum of its inputs and produces a nonlinear \emph{activation function} of that sum. The connectivity between layers can vary, but in general, it has feed-forward connections between layers. These connections each have an associated numerical value, known as a \emph{weight}, that indicates the connection strength. To discern correctness of a given network's predictions in a supervised environment, a \emph{cost function} computes how wrong a prediction is compared to the truth. The partial derivatives of each weight with respect to the cost are used to 
update network parameters through backpropagation, ultimately minimizing the cost function. 



Traditionally 32-bit floating point arithmetic is used for DNN inference. However, the IEEE standard floating point representation is designed for a very broad dynamic range; even 32-bit floating point numbers have a huge dynamic range of over 80 orders of magnitude, far larger than needed for DNNs. While very small values can be important, very large values are not, therefore the design of the numbers creates low information-per-bit based on Shannon maximum entropy \cite{shannon1948mathematical}. Attempts to address this by crafting a fixed-point representation for DNN weights quickly runs up against the \emph{quantization error}. 

The 16-bit (half-precision) form of IEEE floating point, used by Nvidia's accelerators for DNN, reveals the shortcomings of the format: complicated exception cases, gradual underflow, prolific NaN bit patterns, and redundant representations of zero. It is not the representation to design from first principles for a DNN workload. A more recent format, \emph{posit arithmetic}, provides a natural fit to the demands of DNNs both for training and inference.

\subsection{Posit Number System}
The posit number system, a Type III unum, was proposed to improve upon many of the shortcomings of IEEE-754 floating-point arithmetic and to address complaints about the costs of managing the variable size of Type I unums \cite{gustafson2017beating}. (Type II unums are also of fixed size, but require look-up tables that limits their precision \cite{tichy2016unums}.) Posit format provides better dynamic range, accuracy, and consistency between machines than floating point. A posit number is parametrized by $n$, the total number of bits, and $es$, the number of exponent bits. The primary difference between a posit and floating-point number representation is the posit \emph{regime} field, which has a dynamic width like that of a unary number; the regime is a run-length encoded signed value that can be interpreted as in Table \ref{tab:regime}.

\begin{table}[H]
  \caption{Regime Interpretation}
  \label{tab:regime}
  \centering
  \begin{tabular}{c|c|c|c|c|c|c}
    Binary       & \tt{0001} & \tt{001} & \tt{01} & \tt{10} & \tt{110} & \tt{1110} \\
    \hline
    Regime ($k$) &   $-3$ &  $-2$ & $-1$ &  $0$ &   $1$ &    $2$ \\
  \end{tabular}
\end{table}

Two values in the system are reserved: $\tt{10}...\tt{0}$ represents ``Not a Real'' which includes infinity and all other exception cases like $0/0$ and $\sqrt{-1}$, and $\tt{00}...\tt{0}$ represents zero. The full binary representation of a posit number is shown in \eqref{eq:posit}.

\begin{equation}\label{eq:posit}
  \underbrace{
    \xoverbrace{s}                      ^\textup{Sign}
    \xoverbrace{r~r~...~r~\bar{r}~}     ^\textup{Regime}
    \xoverbrace{e_1~e_2~e_3~...~e_{es}~}^\textup{Exponent, if any}
    \xoverbrace{f_1~f_2~f_3~...}        ^\textup{Mantissa, if any}
  }_{n~\textup{Bits}}
\end{equation}

\noindent For a positive posit in such format, the numerical value it represents is given by \eqref{eq:posit_value}

\begin{equation}\label{eq:posit_value}
  (-1) ^ s \times \left( 2 ^ {2 ^ {es}} \right) ^ k \times 2 ^ e \times 1.f
\end{equation}

\noindent where $k$ is the regime value, $e$ is the unsigned exponent (if $es > 0$), and $f$ comprises the remaining bits of the number. If the posit is negative, the 2's complement is taken before using the above decoding. See \cite{gustafson2017beating} for more detailed and complete information on the posit format.

\section{Methodology}


\subsection{Exact Multiply-and-Accumulate (EMAC)}


The fundamental computation within a DNN is the multiply-and-accumulate (MAC) operation. Each neuron within a network is equivalent to a MAC unit in that it performs a weighted sum of its inputs. This operation is ubiquitous across many DNN implementations, however, the operation is usually inexact, \textit{i.e.} limited precision, truncation, or premature rounding in the underlying hardware yields inaccurate results. The EMAC performs the same computation but allocates sufficient padding for digital signals to emulate arbitrary precision. Rounding or truncation within an EMAC unit is delayed until every product has been accumulated, thus producing a result with minimal local error. This minimization of error is especially important when EMAC units are coupled with low-precision data.

In all EMAC units we implement, a number's format is arbitrary as its representation is ultimately converted to fixed-point, which allows for natural accumulation. Given the constraint of low-precision data, we propose to use a variant of the Kulisch accumulator \cite{kulisch2013computer}. In this architecture, a wide register accumulates fixed-point values shifted by an exponential parameter, if applicable, and delays rounding to a post-summation stage. The width of such an accumulator for $k$ multiplications can be computed using \eqref{eq:kulisch_width}
\begin{equation}\label{eq:kulisch_width}
	w_a = \lceil \log_2(k) \rceil + 2 \times \left\lceil \log_2 \left(\frac{max}{min} \right) \right\rceil + 2
\end{equation}
\noindent where $max$ and $min$ are the maximum and minimum values for a number format, respectively. To improve the maximum operating frequency via pipelining, a D flip-flop separates the multiplication and accumulation stages. The architecture easily allows for the incorporation of a bias term -- the accumulator D flip-flop can be reset to the fixed-point representation of the bias so products accumulate on top of it. To further improve accuracy, the \textit{round to nearest} and \textit{round half to even} scheme is employed for the floating point and posit formats. This is the recommended IEEE-754 rounding method and the posit standard.

\subsection{Fixed-point EMAC}
\begin{figure}[H]
  \centering
  \includegraphics[width=0.75\linewidth]{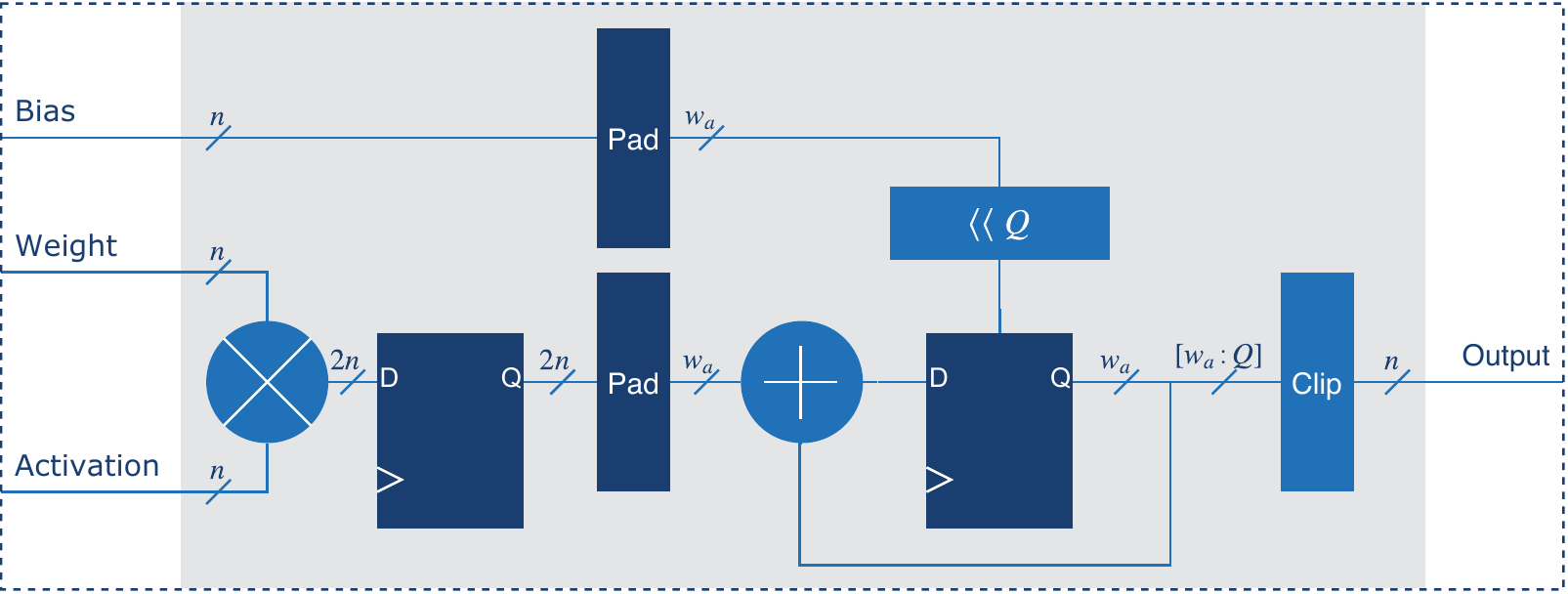}
  \caption{A precision-adaptable (DNN weights and activation) FPGA soft core for fixed-point exact multiply-and-accumulate operation.}
  \label{fig:fixed_mac}
\end{figure}

 The fixed-point EMAC, shown in Fig. \ref{fig:fixed_mac}, accumulates the products of $k$ multiplications and allocates a sufficient range of bits to compute the exact result before truncation. A weight, bias, and activation, each with $q$ fraction bits and $n-q$ integer bits, are the unit inputs. The unnormalized multiplicative result is kept as $2n$ bits to preserve exact precision. The products are accumulated over $k$ clock cycles with the integer adder and D flip-flop combination. The sum of products is then shifted right by $q$ bits and truncated to $n$ bits, ensuring to clip at the maximum magnitude if applicable.

\begin{figure}[H]
  \centering
  \includegraphics[width=\linewidth]{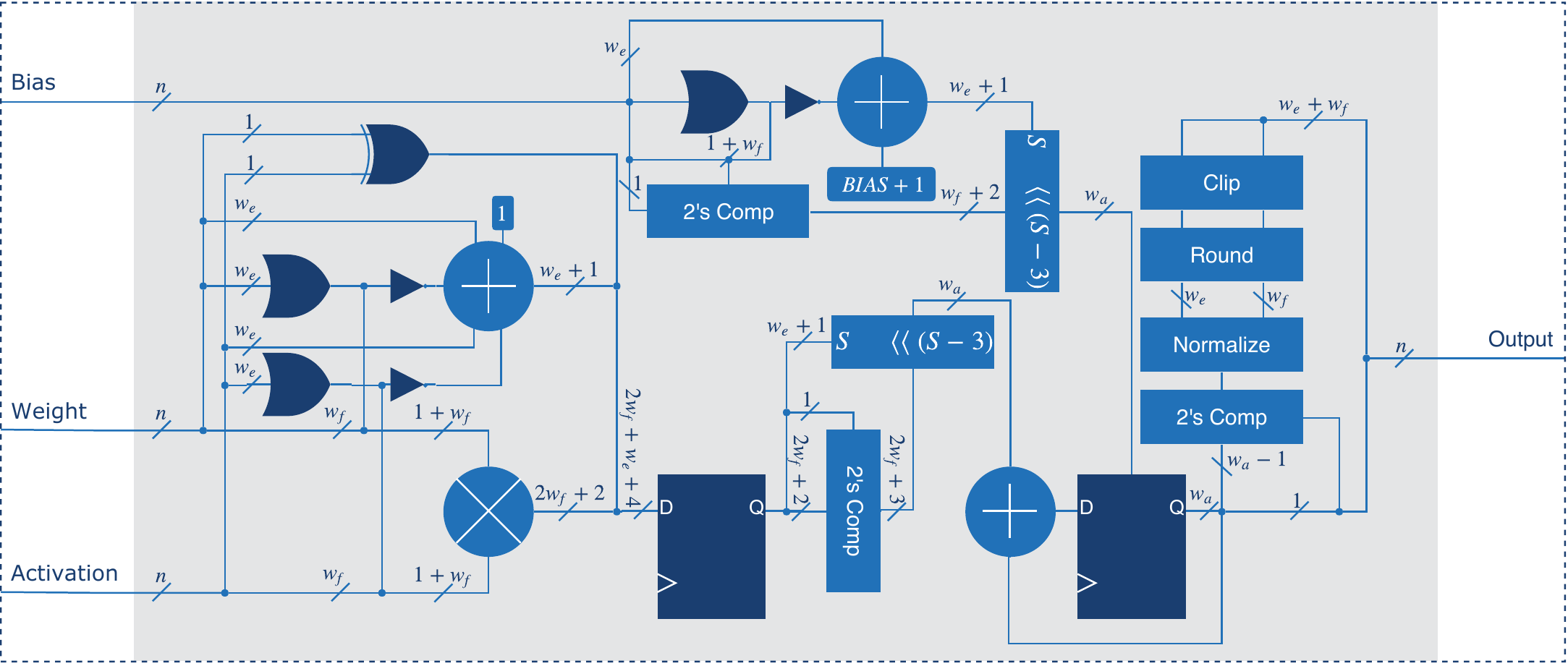}
  \caption{A precision-adaptable (DNN weights and activation) FPGA soft core for the floating-point exact multiply-and-accumulate operation.}
  \label{fig:float_mac}
\end{figure}

\subsection{Floating Point EMAC}

The floating point EMAC, shown in Fig. \ref{fig:float_mac}, also computes the EMAC operation for $k$ pairs of inputs. We do not consider ``Not a Number'' or the ``$\pm$ Infinity'' as inputs don't have these values and the EMAC does not overflow to infinity. Notably, it uses a fixed-point conversion before accumulation to preserve precision of the result. Inputs to the EMAC have a single signed bit, $w_e$ exponent bits, and $w_f$ fraction bits.
Subnormal detection at the inputs appropriately sets the hidden bits and adjusts the exponent. This EMAC scales exponentially with $w_e$ as it's the dominant parameter in computing $w_a$. To convert floating point products to a fixed-point representation, mantissas are converted to 2's complement based on the sign of the product and shifted to the appropriate location in the register based on the product exponent.
After accumulation, inverse 2's complement is applied based on the sign of the sum. If the result is detected to be subnormal, the exponent is accordingly set to `0'. The extracted value from the accumulator is clipped at the maximum magnitude if applicable.

The relevant characteristics of a float number are computed as follows.

\vspace{-5mm}
\begin{align*}
bias &= 2 ^ {w_e - 1} - 1\\
exp_{max} &= 2 ^ {w_e} - 2\\
max &= 2 ^ {exp_{max} - bias} \times (2 - 2 ^ {-w_f})\\
min &= 2 ^ {1 - bias} \times 2 ^ {-w_f}
\end{align*}

\begin{figure*}
  \centering
  \includegraphics[width=0.62\linewidth]{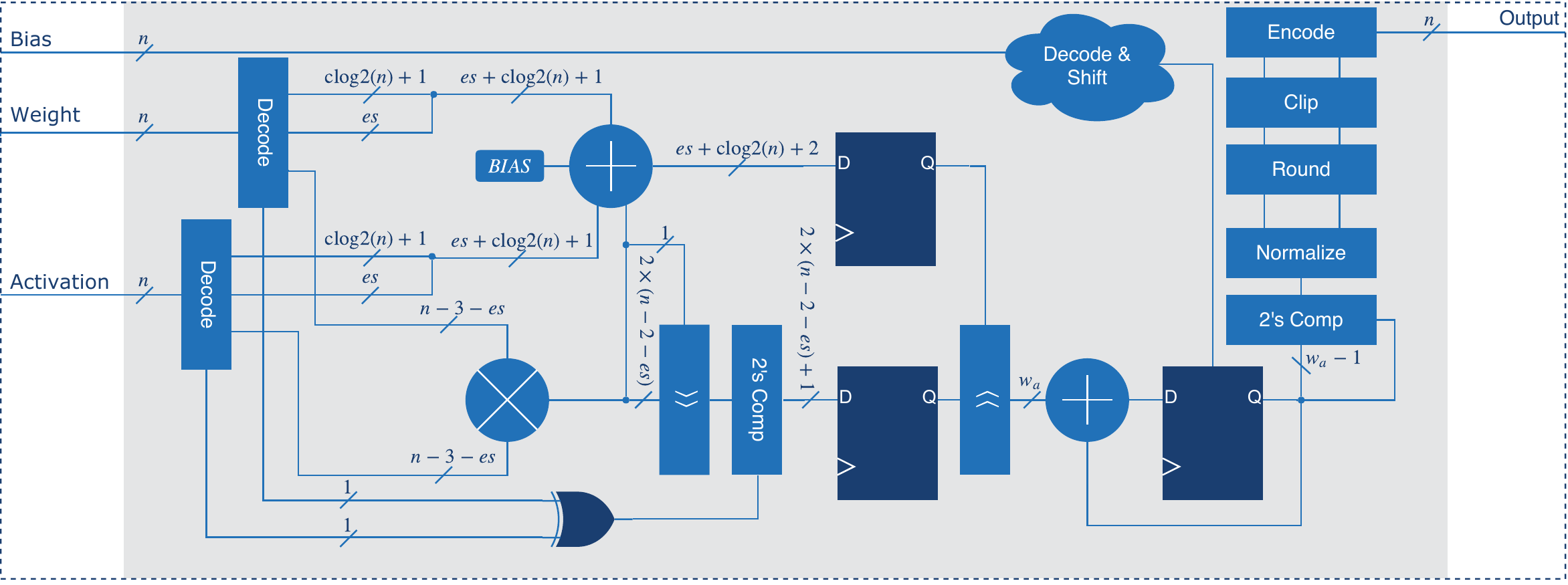}
  \caption{A precision-adaptable (DNN weights and activation) FPGA soft core for the posit exact multiply-and-accumulate operation.}
  \label{fig:posit_mac}
  \vspace{-5mm}
\end{figure*}

\begin{breakablealgorithm}
  \caption{Posit data extraction of $n$-bit input with $es$ exponent bits}\label{alg:posit_data_extract}
  \begin{algorithmic}[1]
  	\begingroup
      \small
      \setlength{\thinmuskip}{2mu}
      \setlength{\medmuskip}{3mu plus 1.5mu minus 3mu}
      \setlength{\thickmuskip}{3.5mu plus 3.5mu}
      \Procedure{Decode}{$\tt{in}$}\Comment{Data extraction of $\tt{in}$}
        \State ${\tt{nzero}} \gets |{\tt{in}}$\Comment{'1' if $\tt{in}$ is nonzero}
        \State ${\tt{sign}} \gets {\tt{in}}[n\unaryminus1]$\Comment{Extract sign}
        \State ${\tt{twos}} \gets (\{n\unaryminus1\{{\tt{sign}}\}\}\oplus {\tt{in}}[n\unaryminus2:0])+{\tt{sign}}$\Comment{2's Comp.}
        \State ${\tt{rc}} \gets {\tt{twos}}[n\unaryminus2]$\Comment{Regime check}
        \State ${\tt{inv}} \gets \{n\unaryminus1\{{\tt{rc}}\}\} \oplus {\tt{twos}}$\Comment{Invert 2's}
        \State ${\tt{zc}} \gets \textup{LZD}({\tt{inv}})$\Comment{Count leading zeros}
        \State ${\tt{tmp}} \gets {\tt{twos}}[n\unaryminus4:0] \ll ({\tt{zc}}-1)$\Comment{Shift out regime}
        \State ${\tt{frac}} \gets \{{\tt{nzero}}, {\tt{tmp}}[n\unaryminus es \unaryminus 4:0]\}$\Comment{Extract fraction}
        \State ${\tt{exp}} \gets {\tt{tmp}}[n\unaryminus 4:n\unaryminus es \unaryminus 3]$\Comment{Extract exponent}
        \State ${\tt{reg}} \gets {\tt{rc}}~?~{\tt{zc}}\unaryminus1:\unaryminus {\tt{zc}}$\Comment{Select regime}
        \State \textbf{return} $\tt{sign}$, $\tt{reg}$, $\tt{exp}$, $\tt{frac}$
      \EndProcedure
    \endgroup
  \end{algorithmic}
\end{breakablealgorithm}

\subsection{Posit EMAC}

The posit EMAC, detailed in Fig. \ref{fig:posit_mac}, computes the operation of $k$ pairs of inputs. We do not consider ``Not a Real'' in this implementation as all inputs are expected to be real numbers and posits never overflow to infinity. Inputs to the EMAC are decoded in order to extract the sign, regime, exponent, and fraction. As the regime bit field is of dynamic size, this process is nontrivial. Algorithm \ref{alg:posit_data_extract} describes the data extraction process. To mitigate needing both a leading ones detector (LOD) and leading zeros detector (LZD), we invert the two's complement of the input (line 5) so that the regime always begins with a `0'. The regime is accordingly adjusted using the regime check bit (line 11). After decoding inputs, multiplication and converting to fixed-point is performed similarly to that of floating point. Products are accumulated in a register, or quire in the posit literature, of width $qsize$ as given by \eqref{eq:quire_width}.
\begin{align}\label{eq:quire_width}
\vspace{-2mm}
qsize &= 2 ^ {es+2} \times (n-2)  + 2 + \lceil \log_2(k) \rceil,~n\ge3
\end{align}

To avoid using multiple shifters in fixed-point conversion, the scale factor ${\tt{sf_{mult}}}$ is biased by $bias=2 ^ {es+1} \times (n-2)$ such that its minimum value becomes 0. After accumulation, the scale factor is unbiased by $bias$ before entering the convergent rounding and encoding stage. Algorithm \ref{alg:posit_edp} gives the procedure for carrying out these operations.

\begin{breakablealgorithm}
  \caption{Posit EMAC operation for $n$-bit inputs each with $es$ exponent bits}\label{alg:posit_edp}
  \begin{algorithmic}[1]
  	\begingroup
      \small
      \setlength{\thinmuskip}{2mu}
      \setlength{\medmuskip}{3mu plus 1.5mu minus 3mu}
      \setlength{\thickmuskip}{3.5mu plus 3.5mu}
      \Procedure{PositEMAC}{$\tt{weight,activation}$}
        \State ${\tt{sign_w}}, {\tt{reg_w}}, {\tt{exp_w}}, {\tt{frac_w}} \gets \text{\headersty{Decode}}({\tt{weight}})$
        \State ${\tt{sign_a}}, {\tt{reg_a}}, {\tt{exp_a}}, {\tt{frac_a}} \gets \text{\headersty{Decode}}({\tt{activation}})$
        \State ${\tt{sf_w}} \gets \{{\tt{reg_w}}, {\tt{exp_w}}\}$\Comment{Gather scale factors}
        \State ${\tt{sf_a}} \gets \{{\tt{reg_a}}, {\tt{exp_a}}\}$
        \algrule \hspace{-1.75mm}\textbf{Multiplication}
        \State ${\tt{sign_{mult}}} \gets {\tt{sign_w}} \oplus {\tt{sign_a}}$
        \State ${\tt{frac_{mult}}} \gets {\tt{frac_w}} \times {\tt{frac_a}}$
        \State ${\tt{ovf_{mult}}} \gets {\tt{frac_{mult}}}[{\tt{MSB}}]$\Comment{Adjust for overflow}
        \State ${\tt{normfrac_{mult}}} \gets {\tt{frac_{mult}}} \gg {\tt{ovf_{mult}}}$
        \State ${\tt{sf_{mult}}} \gets {\tt{sf_{w}}} + {\tt{sf_{a}}} + {\tt{ovf_{mult}}}$
        \algrule \hspace{-1.75mm}\textbf{Accumulation}
        \State ${\tt{fracs_{mult}}} \gets {\tt{sign_{mult}}}~?~{\unaryminus\tt{frac_{mult}}}:{\tt{frac_{mult}}}$
        \State ${\tt{sf_{biased}}} \gets {\tt{sf_{mult}}} + bias$\Comment{Bias the scale factor}
        \State ${\tt{fracs_{fixed}}} \gets {\tt{fracs_{mult}}} \ll {\tt{sf_{biased}}}$\Comment{Shift to fixed}
        \State ${\tt{sum_{quire}}} \gets {\tt{fracs_{fixed}}} + {\tt{sum_{quire}}}$\Comment{Accumulate}
        \algrule \hspace{-1.75mm}\textbf{Fraction \& SF Extraction}
        \State ${\tt{sign_{quire}}} \gets {\tt{sum_{quire}}}[{\tt{MSB}}]$
        \State ${\tt{mag_{quire}}} \gets {\tt{sign_{quire}}}~?~{\unaryminus\tt{sum_{quire}}}:{\tt{sum_{quire}}}$
        \State ${\tt{zc}} \gets \textup{LZD}({\tt{mag_{quire}}})$
        \State ${\tt{frac_{quire}}} \gets {\tt{mag_{quire}}}[2{\times}(n\unaryminus 2\unaryminus es)\unaryminus 1{+}{\tt{zc}}:{\tt{zc}}]$
        \State ${\tt{sf_{quire}}} \gets {\tt{zc}} \unaryminus bias$
        \algrule \hspace{-1.75mm}\textbf{Convergent Rounding \& Encoding}
        \State ${\tt{nzero}} \gets |{\tt{frac_{quire}}}$
        \State ${\tt{sign_{sf}}} \gets {\tt{sf_{quire}}}[{\tt{MSB}}]$
    	\State ${\tt{exp}} \gets {\tt{sf_{quire}}}[es\unaryminus 1:0]$\Comment{Unpack scale factor}
        \State ${\tt{reg_{tmp}}} \gets {\tt{sf_{quire}}}[{\tt{MSB}}\unaryminus1:es]$
        \State ${\tt{reg}} \gets {\tt{sign_{sf}}}~?~\unaryminus{\tt{reg_{tmp}}}:{\tt{reg_{tmp}}}$
        \State ${\tt{ovf_{reg}}} \gets {\tt{reg}}[{\tt{MSB}}]$\Comment{Check for overflow}
        \State ${\tt{reg_f}} \gets {\tt{ovf_{reg}}}~?~\{\{\lceil\log_2(n)\rceil\unaryminus 2 \{{\tt{1}}\}\}), {\tt{0}}\} : {\tt{reg}}$
        \State ${\tt{exp_f}} \gets ({\tt{ovf_{reg}}}|{\sim}{\tt{nzero}}|{(\tt{\&{\tt{reg_f}}}}))~?~\{es\{{\tt{0}}\}\}:{\tt{exp}}$
        \State ${\tt{tmp1}} \gets \{{\tt{nzero}}, {\tt{0}}, {\tt{exp_f}}, {\tt{frac_{quire}}}[{\tt{MSB}}\unaryminus 1:0],$ \par
        $~~~~~~~~~~\{n\unaryminus 1 \{{\tt{0}}\}\}\}$
        \State ${\tt{tmp2}} \gets \{{\tt{0}}, {\tt{nzero}}, {\tt{exp_f}}, {\tt{frac_{quire}}}[{\tt{MSB}}\unaryminus 1:0],$ \par
        $~~~~~~~~~~\{n\unaryminus 1 \{{\tt{0}}\}\}\}$
        \State ${\tt{ovf_{regf}}} \gets \&{\tt{reg_f}}$
        \If {${\tt{ovf_{regf}}}$}
        	\State ${\tt{shift_{neg}}} \gets {\tt{reg_{f}}} - 2$
			\State ${\tt{shift_{pos}}} \gets {\tt{reg_{f}}} - 1$
		\Else
			\State ${\tt{shift_{neg}}} \gets {\tt{reg_{f}}} - 1$
			\State ${\tt{shift_{pos}}} \gets {\tt{reg_{f}}}$
		\EndIf
        \State ${\tt{tmp}} \gets {\tt{sign_{sf}}}~?~{\tt{tmp2}} \gg {\tt{shift_{neg}}} : {\tt{tmp1}} \gg {\tt{shift_{pos}}}$
        \State ${\tt{lsb}}, {\tt{guard}} \gets {\tt{tmp}}[{\tt{MSB}}\unaryminus (n\unaryminus 2):{\tt{MSB}}\unaryminus (n\unaryminus 1)]$
        \State ${\tt{round}} \gets {\sim}({\tt{ovf_{reg}}}|{\tt{ovf_{regf}}})~? $ \par
        $~~~~~~~~~~~~~~~~~~(~{\tt{guard}}~\&~({\tt{lsb}}~|~(|{\tt{tmp}}[{\tt{MSB}}\unaryminus n : 0]))~) : {\tt{0}}$
        \State ${\tt{result_{tmp}}} \gets {\tt{tmp}}[{\tt{MSB}} : {\tt{MSB}}\unaryminus n{+}1] {+} {\tt{round}}$
		\State ${\tt{result}} \gets {\tt{sign_{quire}}}~?~\unaryminus{\tt{result_{tmp}}} : {\tt{result_{tmp}}}$
        \State \textbf{return} ${\tt{result}}$
      \EndProcedure
    \endgroup
  \end{algorithmic}
\end{breakablealgorithm}

The relevant characteristics of a posit number are computed as follows.
\vspace{-2mm}
\begin{align*}
useed  &= 2^{2^{es}} \\
max &= useed^{n-2} \\
min &= useed^{-n+2}
\end{align*}



\subsection{Deep Positron}

We assemble a custom DNN architecture that is parametrized by data width, data type, and DNN hyperparameters (\textit{e.g.} number of layers, neurons per layer, \textit{etc.}), as shown in Fig.~\ref{fig:deep_positron_arch}. Each layer contains dedicated EMAC units with local memory blocks for weights and biases. Storing DNN parameters in this manner minimizes latency by avoiding off-chip memory accesses. The compute cycle of each layer is triggered when its directly preceding layer has terminated computation for an input. This flow performs inference in a parallel streaming fashion. The ReLU activation is used throughout the network, except for the affine readout layer. A main control unit controls the flow of input data and activations throughout the network using a finite state machine.

\section{Experimental Results}

\subsection{EMAC Analysis and Comparison}
We compare the hardware implications across three numerical format parameters that each EMAC has on a Virtex-7 FPGA (xc7vx485t-2ffg1761c). Synthesis results are obtained through Vivado 2017.2 and optimized for latency by targeting the on-chip DSP48 slices. Our preliminary results indicate that the posit EMAC is competitive with the floating point EMAC in terms of energy and latency.
At lower values of $n\le7$, the posit number system has higher dynamic range as emphasized by \cite{gustafson2017beating}. We compute dynamic range as $\log_{10}\left( \frac{max}{min} \right)$. While neither the floating point or posit EMACs can compete with the energy-delay-product (EDP) of fixed-point, they both are able to offer significantly higher dynamic range for the same values of $n$. Furthermore, the EDPs of the floating point and posit EMACs are similar.


\vspace{-4mm}
\begin{figure}[H]
  \centering
  \includegraphics[width=0.6\linewidth]{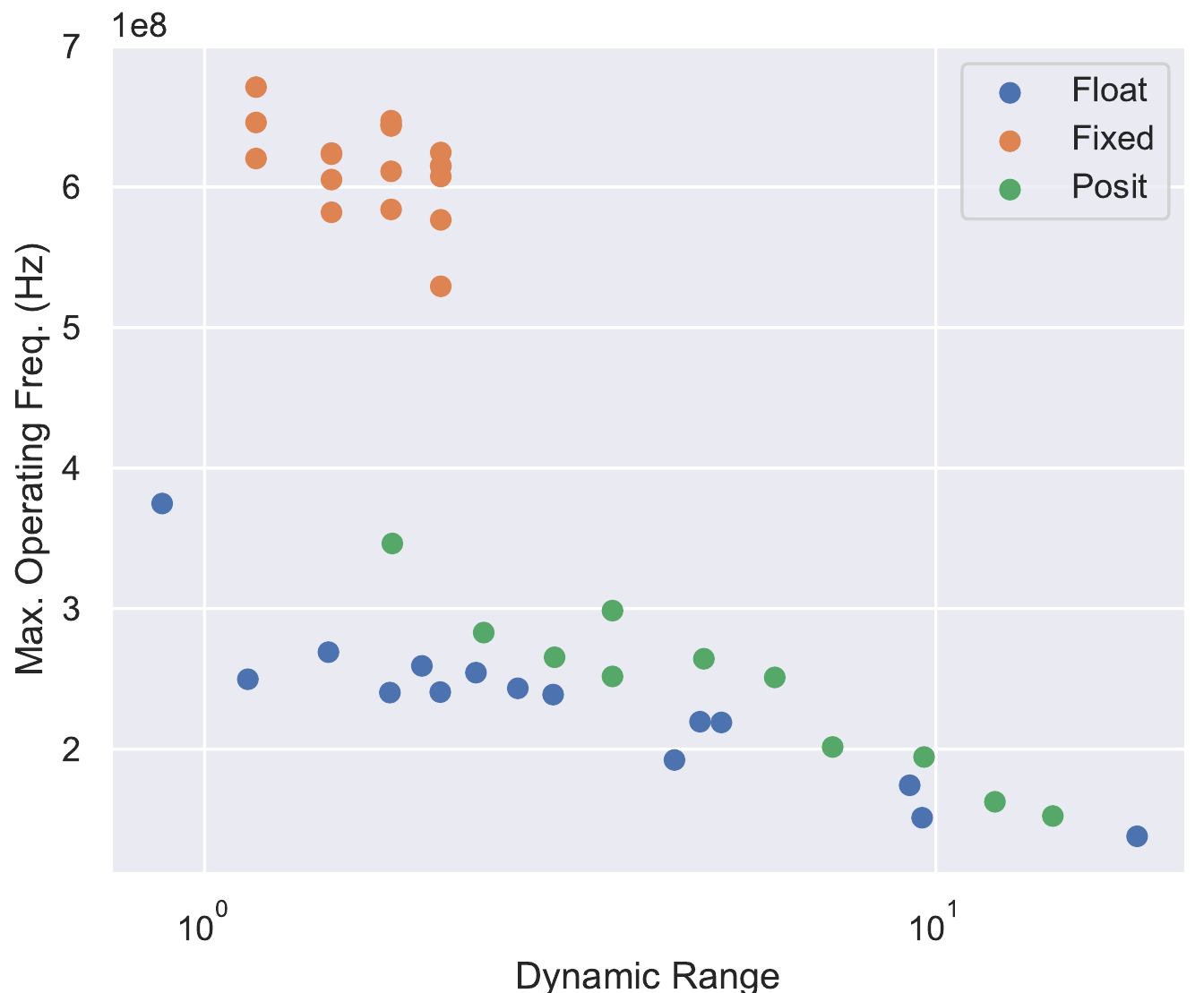}
  \caption{Dynamic Range vs. Maximum Operating Frequency (Hz) for the EMACs implemented on Xilinx Virtex-7 FPGA.}
  \label{fig:dr_vs_mof}
  \vspace{-4mm}
\end{figure}

\vspace{-4mm}
\begin{figure}[H]
  \centering
  \includegraphics[width=0.6\linewidth]{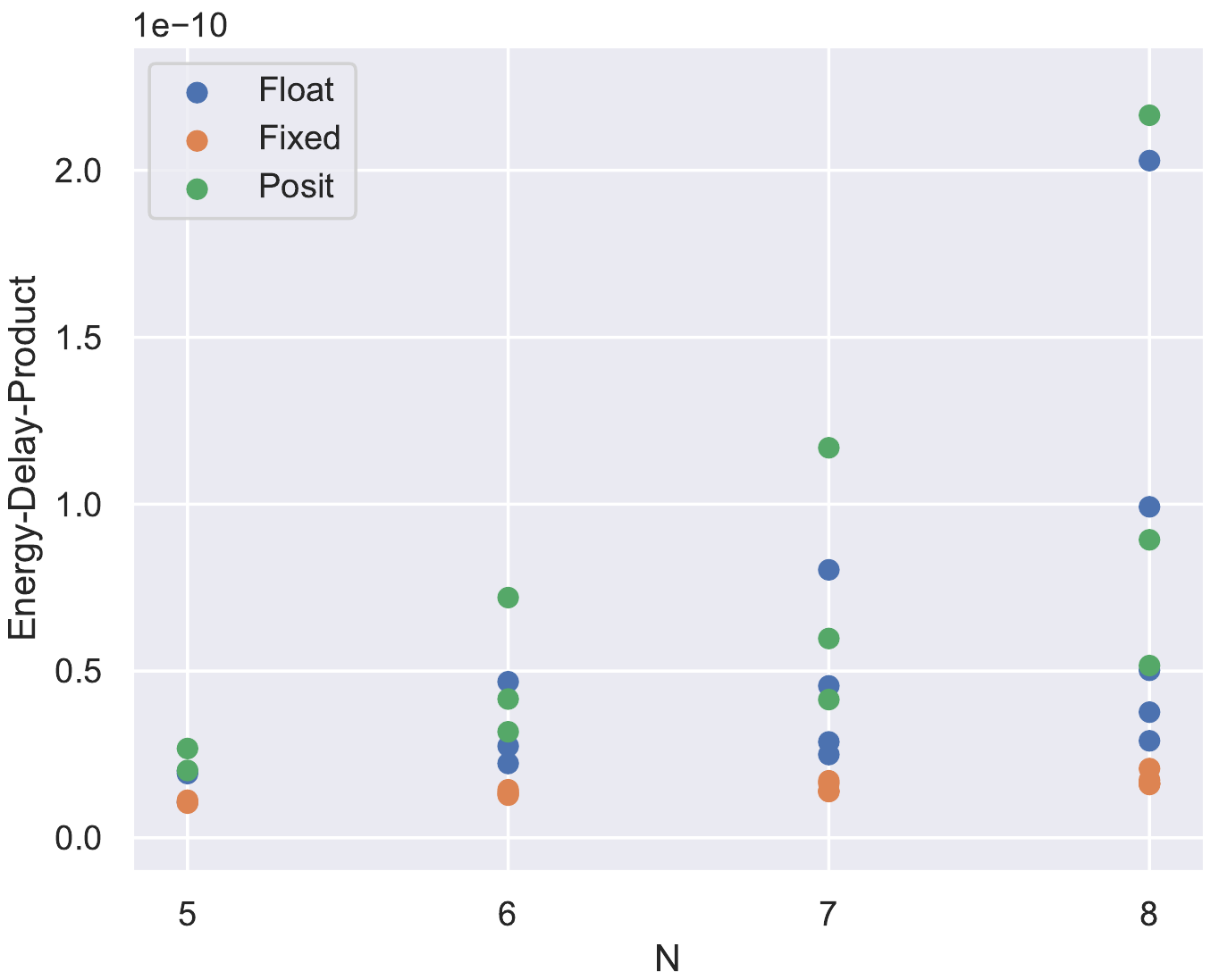}
  \vspace{-2mm}
  \caption{$n$ vs. energy-delay-product for the EMACs implemented on Xilinx Virtex-7 FPGA.}
  \label{fig:n_vs_edp}
  \vspace{-4mm}
\end{figure}

Fig. \ref{fig:dr_vs_mof} shows the synthesis results for the dynamic range of each format against maximum operating frequency. As expected, the fixed-point EMAC achieves the lowest datapath latencies as it has no exponential parameter, thus a narrower accumulator. In general, the posit EMAC can operate at a higher frequency for a given dynamic range than the floating point EMAC. Fig. \ref{fig:n_vs_edp} shows the EDP across different bit-widths and as expected fixed-point outperforms for all bit-widths.

\begin{figure}[H]
  \centering
  \includegraphics[width=0.6\linewidth]{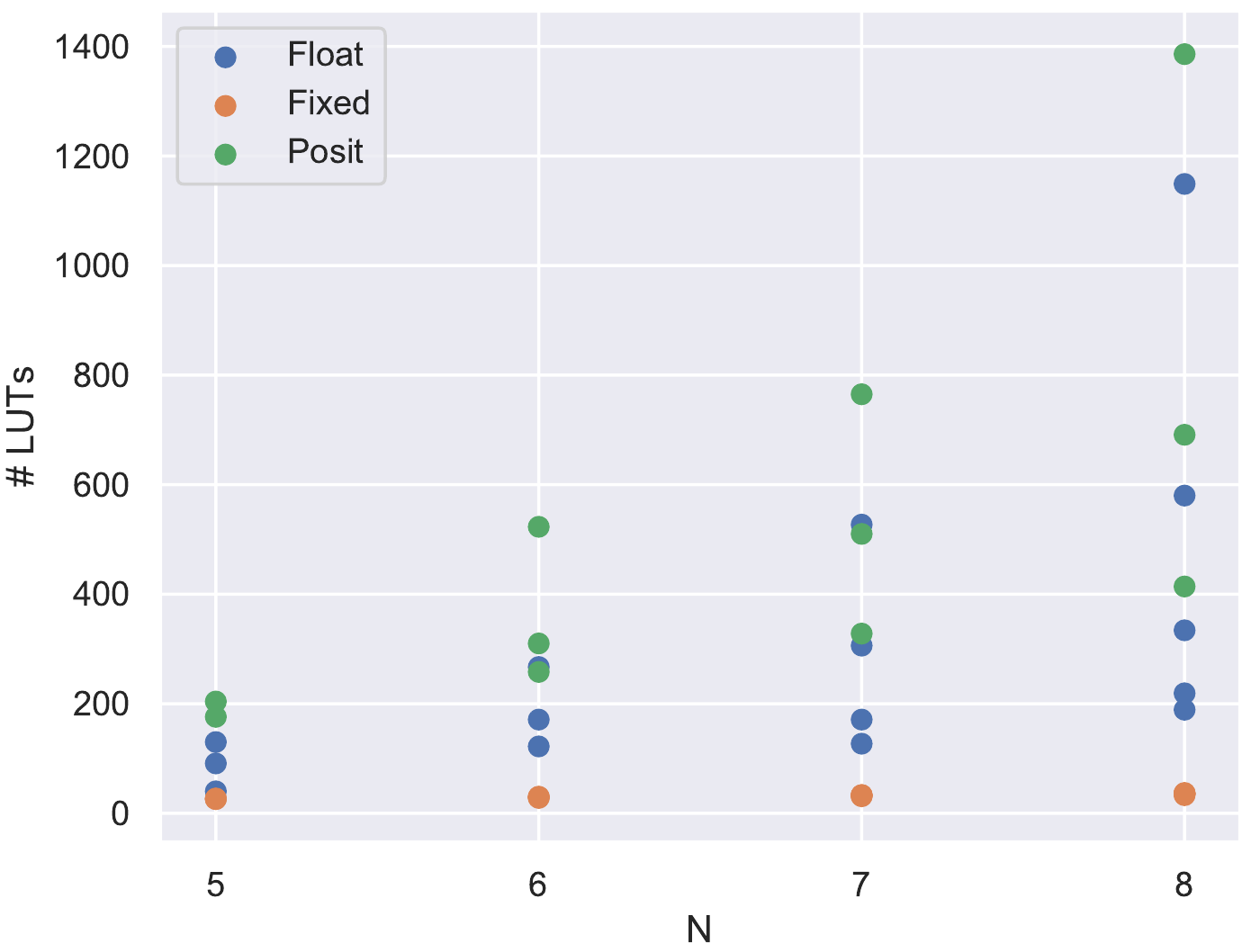}
  \vspace{-2mm}
  \caption{$n$ vs. LUT Utilization for the EMACs implemented on Xilinx Virtex-7 FPGA.}
  \label{fig:n_vs_lut}
 \vspace{-3mm}
\end{figure}

The LUT utilization results against numerical precision $n$ are shown in Fig. \ref{fig:n_vs_lut}, where posit generally consumes a higher amount of resources. This limitation can be attributed to the more involved decoding and encoding of inputs and outputs.

\vspace{-4mm}

\subsection{Deep Positron Performance}
We compare the performance of Deep Positron on three datasets and all possible combinations of [5,8] bit-widths for the three numerical formats. Posit performs uniformly well across all the three datasets for 8-bit precision, shown in Table~\ref{table:1-1}, and has similar accuracy as fixed-point and float in sub 8-bit. Best results are when posit has $es \in \{0, 2\}$ and floating point has $w_e \in \{3, 4\}$. As expected, the best performance drops sub 8-bit by [0-4.21]\% compared to 32-bit floating-point. In all experiments, the posit format either outperforms or matches the performance of floating and fixed-point. Additionally, with 24 fewer bits, posit matches the performance of 32-bit floating point on the Iris classification task.

\begin{figure}[H]
  \centering
  \includegraphics[width=0.6\linewidth]{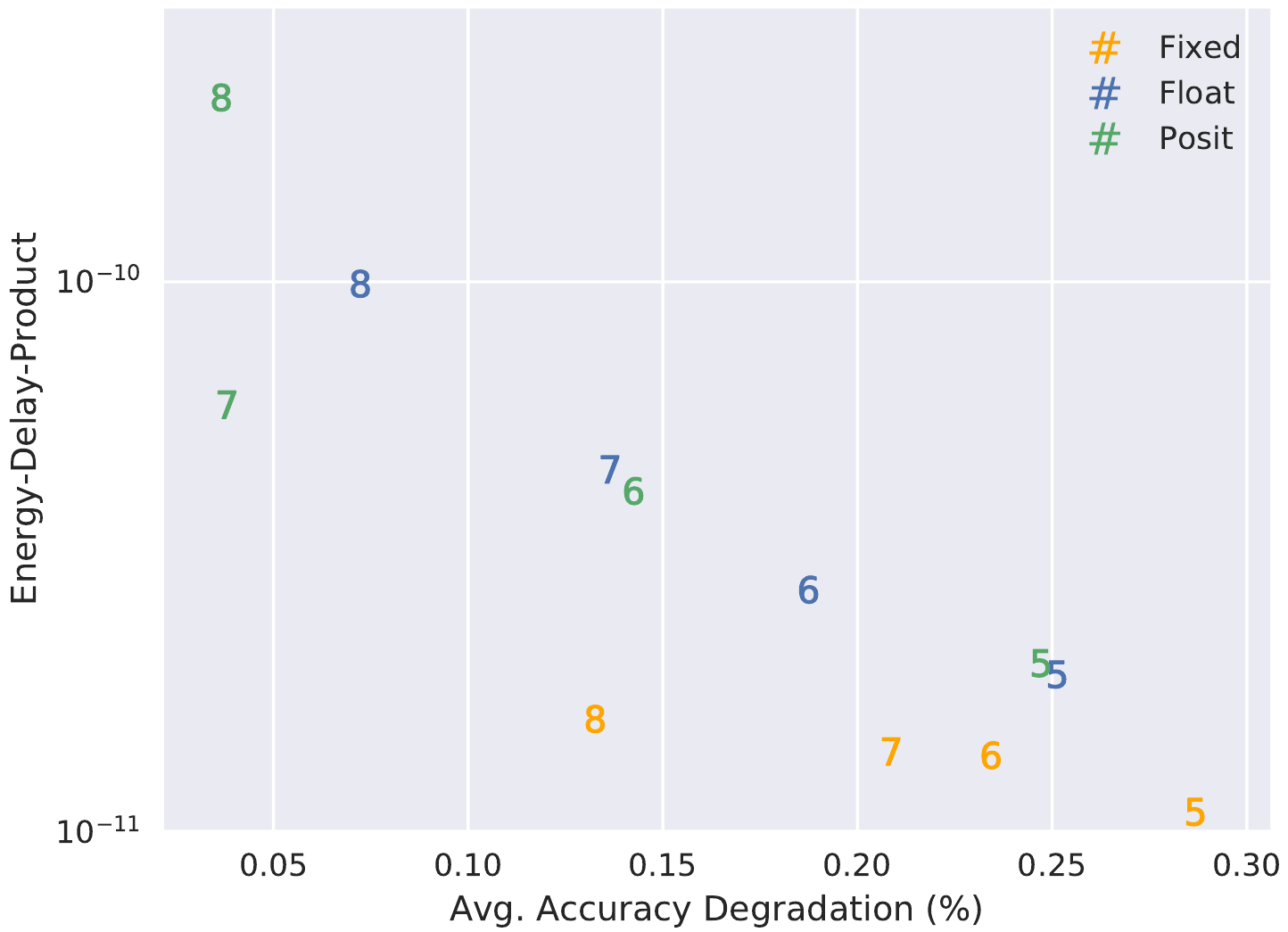}
  \vspace{-2mm}
  \caption{Average accuracy degradation vs. energy-delay-product for the EMACs implemented on Xilinx Virtex-7 FPGA. Numbers correspond with the bit-width of a numerical format.}
  \label{fig:deg_vs_edp}
  \vspace{-3mm}
\end{figure}

Fig. \ref{fig:deg_vs_edp} shows the lowest accuracy degradation per bit width against EDP. The results indicate posits achieve better performance compared to the floating and fixed-point formats at a moderate cost.

\vspace{-4mm}

\begin{table}[ht!]
\caption{Deep Positron performance on low-dimensional datasets with 8-bit EMACs.}
\centering
\resizebox{0.48\textwidth}{!}{
\begin{tabular}{|c|c|c|c|c|c|} 
\hline
 \rowcolor[gray]{.9}
  & & & Accuracy & &\\
  \hline
  \rowcolor[gray]{.9}
   Dataset & Inference size & Posit & Floating-point & Fixed-point & 32-bit Float\\
    \hline 
    Wisconsin Breast Cancer~\cite{WBC} & 190 & 85.89\% & 77.4\% & 57.8\% & 90.1\% \\
    \hline
     Iris~\cite{IRIS} & 50 &  98\% & 96\% & 92\% & 98\%\\
    \hline
     Mushroom~\cite{Mashrom}& 2708 & 96.4\% & 96.4\% & 95.9\% & 96.8\%\\
    \hline
\end{tabular}
}
\label{table:1-1}
\vspace{-3.9mm}
\end{table}
\section{Related Work}
Research on low-precision arithmetic for neural networks dates to circa 1990 ~\cite{iwata1989artificial,hammerstrom1990vlsi} using fixed-point and floating point. Recently, several groups have shown that it is possible to perform inference in DNNs with 16-bit fixed-point representations~\cite{Courbariaux14,bengio2013deep}. However, most of these studies compare DNN inference for different bit-widths. Few research teams have performed a comparison with same bit-widths across different number systems coupled with FPGA soft processors. For example, Hashemi \textit{et al.} demonstrate DNN inference with 32-bit fixed-point and 32-bit floating point on the LeNet, ConvNet, and AlexNet DNNs, where the energy consumption is reduced by $\sim$12\% and $<$1\% accuracy loss with fixed-point~\cite{hashemi2017understanding}. Most recently, Chung \textit{et al.} proposed an accelerator, Brainwave, with a spatial 8-bit floating point, called \textit{ms-fp8}. The \textit{ms-fp8} format improves the throughput by 3$\times$ over 8-bit fixed-point on a Stratix-10 FPGA \cite{Microsoft2018}. 
 
This paper also relates to three previous works that use posits in DNNs. The first DNN architecture using the posit number system was proposed by Langroudi \textit{et al.}
\cite{Langroudi2018}. The work demonstrates that, with $<$1\% accuracy degradation, DNN parameters can be represented using 7-bit posits for AlexNet on the ImageNet corpus and that posits require $\sim$30\% less memory utilization for the LeNet, ConvNet, and AlexNet neural networks in comparison to the fixed-point format. Secondly, Cococcioni \textit{et al.} \cite{cococcioni2018} discuss the effectiveness of posit arithmetic for application to autonomous driving. They consider an implementation of the Posit Processing Unit (PPU) as an alternative to the Floating point Processing Unit (FPU) since the self-driving car standards require 16-bit floating point representations for the safety-critical application. Recently, Jeff Johnson proposed a log float format as a combination of the posit format and the logarithmic version of the EMAC operation called the exact log-linear multiply-add (ELMA). This work shows that ImageNet classification using the ResNet-50 DNN architecture can be performed with $<$1\% accuracy degradation \cite{johnson2018rethinking}. It also shows that 4\% and 41\% power consumption reduction can be achieved by using an 8/38-bit ELMA in place of an 8/32-bit integer multiply-add and an IEEE-754 float16 fused multiply-add, respectively.

This paper is inspired by the earlier studies and demonstrates that posit arithmetic with ultra-low precision ($\leq$8-bit) is a natural choice for DNNs performing low-dimensional tasks. A precision-adaptable, parameterized FPGA soft core is used for comprehensive analysis on the Deep Positron architecture with same bit-width for fixed, floating-point, and posit formats.
\section{Conclusions}




In this paper, we show that the posit format is well suited for deep neural networks at ultra-low precision ($\leq$8-bit). We show that precision-adaptable, reconfigurable exact multiply-and-accumulate designs embedded in a DNN are efficient for inference. Accuracy-sensitivity studies for Deep Positron show robustness at 7-bit and 8-bit widths. In the future, the success of DNNs in real-world applications will equally rely on the underlying platforms and architectures as much as the algorithms and data. Full-scale DNN accelerators with low-precision posit arithmetic will play an important role in this domain.
\vspace{-3mm}
\bibliographystyle{IEEEtran}
\bibliography{DATE}

\end{document}